\documentclass[10pt,conference,compsocconf]{IEEEtran}
\usepackage{times}

\usepackage{caption}
\usepackage[noadjust]{cite}
\usepackage{amsmath,amssymb,amsfonts}
\usepackage{algorithmic}
\usepackage{graphicx}
\usepackage{textcomp}
\usepackage{xcolor,subfigure}
\usepackage{dblfloatfix}  
\usepackage{float}
\ifCLASSINFOpdf
\else
\fi
\begin{document}
\pagenumbering{gobble}
%
\title{\textbf{\Large Adversarial Training for Deep Learning-based  \\[-1.5ex] Intrusion Detection Systems}\\[0.2ex]}


	\author{\IEEEauthorblockN{~\\[-0.4ex]\large Islam Debicha\\[0.3ex]\normalsize}
	\IEEEauthorblockA{Royal Military Academy \\
		and Université Libre de Bruxelles\\
		Brussels, Belgium\\
		Email: {\tt debichasislam@gmail.com}  }
	\and
	\IEEEauthorblockN{~\\[-0.4ex]\large Thibault Debatty\\[0.3ex]\normalsize}
	\IEEEauthorblockA{Royal Military Academy\\
		Brussels, Belgium\\
		\qquad \qquad	Email: {\tt thibault.debatty@rma.ac.be} \qquad \qquad}
	\and
	\IEEEauthorblockN{~\\[-0.4ex]\large Jean-Michel Dricot\\[0.3ex]\normalsize}
	\IEEEauthorblockA{Université Libre de Bruxelles\\
		Brussels, Belgium\\
		\qquad \qquad \qquad Email: {\tt jdricot@ulb.ac.be}\qquad \qquad \qquad}
	\and
	\IEEEauthorblockN{~\\[-0.4ex]\large Wim Mees\\[0.3ex]\normalsize}
	\IEEEauthorblockA{Royal Military Academy\\
		Brussels, Belgium\\
		Email: {\tt wim.mees@rma.ac.be}  }
	
		}


%


\maketitle

\begin{abstract}
Nowadays, Deep Neural Networks (DNNs) report state-of-the-art results in many machine learning areas, including intrusion detection. Nevertheless, recent studies in computer vision have shown that DNNs can be vulnerable to adversarial attacks that are capable of deceiving them into misclassification by injecting specially crafted data. In security-critical areas, such attacks can cause serious damage; therefore, in this paper, we examine the effect of adversarial attacks on deep learning-based intrusion detection. In addition, we investigate the effectiveness of adversarial training as a defense against such attacks. Experimental results show that with sufficient distortion, adversarial examples are able to mislead the detector and that the use of adversarial training can improve the robustness of intrusion detection. 
\end{abstract}



%
\IEEEpeerreviewmaketitle

\section{Introduction}
With the growth of the computer and telecommunications industry and the expansion of the Internet, there has been a significant escalation of cyberattacks targeting all types of networks, as attackers are increasingly motivated to develop new ways to penetrate systems, given the great reward. As a result, securing these networks has become a crucial area of interest. Intrusion Detection Systems (IDS), which are designed to detect and identify anomalies and attacks, are gaining popularity and are presented as one of the solutions against these cyber-attacks.

There are mainly two types of intrusion detection systems: those based on signatures and those based on anomaly detection. The first one works more or less in the same way as most antivirus systems by maintaining a database with all known attack signatures.  An exhaustive comparison of the incoming traffic with the signature database allows the system to determine if it represents an attack. These systems are remarkably effective at detecting known attacks and offer high accuracy but are obviously unable to detect zero-day exploits. This is essentially what drives the use of anomaly-based intrusion detection systems that work by modeling the normal behavior of traffic and network activities and then comparing new traffic to this baseline. 

 Several research studies have examined the use of different Machine Learning (ML) techniques to improve the accuracy of anomaly-based IDS \cite{kevric2017effective}\cite{debicha2020efficient}. Nevertheless, the lack of transferability and the dependence of traditional machine learning on domain knowledge (feature engineering) have been among the main reasons for substituting them with DNNs which not only solved these problems but also yielded, in most cases, the highest accuracies, making them the state-of-the-art in the field of anomaly-based intrusion detection \cite{wang2018deep}.

Despite their popularity, DNNs have proven to be vulnerable to adversarial attacks in computer vision where, by introducing imperceptible changes in an image, an adversary can mislead the classifier. When applied to machine learning-based security products, these attacks can lead to a critical security breach. Although a considerable amount of studies has been conducted on adversarial attacks in computer vision, there are very few studies on this issue in intrusion detection. Therefore, the contribution of this paper is double: (1) we study the effect of adversarial attacks on deep learning-based intrusion detection systems. For that, three adversarial attacks are tested: Fast Gradient Sign Method (FGSM) \cite{goodfellow2014explaining}, Basic Iterative Method (BIM) \cite{kurakin2016adversarial} and Projected Gradient Descent (PGD) \cite{madry2017towards} showing that adversarial attacks are able, given enough strength, to mislead the IDS significantly. In addition, (2) this is the first study, to the best of our knowledge, to examine the effectiveness of adversarial training as a defense against adversarial attacks for intrusion detection systems.

In what follows, we briefly recall the concept of DNNs, present an overview of related work and explain the idea of adversarial examples in Section \ref{sec:back}. The experimental approach is explained in Section \ref{sec:ea}. Results and discussions are presented in Section \ref{sec:er}. Concluding remarks and suggestions for possible follow-up work are given in Section \ref{sec:cfw}.


\section{Background}
\label{sec:back}

\subsection{Deep Neural Network (DNN)}
DNN refers to a machine learning algorithm made up of multiple interconnected layers where each layer is composed of several nodes - called neurons. Within each neuron, an activation function operates as a basic computing unit. The activation function input on a neuron is the parameter-weighted output of the immediately preceding layer, whilst each layer's output is at the same time the next layer's input.

Frequently described as an end-to-end machine learning process, DNN is capable of learning complex patterns based on limited prior knowledge of input data representation. As a result, deep learning models are widely applied to address large-scale data problems that are frequently inadequately handled by traditional machine learning algorithms. DNN layers fall into three categories: the input layer, the output layer, and, in between, the hidden layer. For large-scale input data, it may be necessary to use several hidden layers so as to learn the subjacent correlation.

DNN can be seen as a function \textit{f}($\cdot$), \textit{f} $\in$ \textit{F}: $\mathbb{R}^n$ $\rightarrow$ $\mathbb{R}^m$. let $\Theta$ be the DNN parameters. Training the model involves finding the optimal parameters $\Theta$ where the loss function \textit{J} (e.g., cross-entropy) is minimal. 

For the classification task, the outputs of the last layer in DNN are called logits. The softmax function is added after the last layer in order to transform these logits into a probability distribution, i.e., $0 \leq y_i \leq 1$ and $y_1$ + . + $y_m = 1$ where $y_i$ is interpreted as the probability that input x has class $i$. The label with the highest probability C(x) = argmax $y_i$ is assigned as the class of the input $x$. 

Let $Z(x) = z$ be the output of all layers excluding Softmax, thus the full DNN is $F(x)$ = softmax$(Z(x)) = y$. 
At the neuron level, the input is first linearly transformed using  weights $\tilde{\theta}$ and baises $\hat{\theta}$ , and then subjected to a non-linear activation function $\sigma$ (e.g., ReLU). The DNN model is a chain function : 
\begin{equation}
F = softmax \circ F_{n} \circ F_{n-1} \circ \cdots \circ F_{1}
\end{equation}
  Where:
\begin{equation}
  F_{i}(x) = \sigma(\tilde{\theta_{i}} . x + \hat{\theta_{i}} ) 
\end{equation}

\subsection{Related work}
\label{sec:rw}

Several studies have shown the effectiveness of the DNN for intrusion detection systems in different types of networks. \cite{diro2018leveraging} has proposed an LSTM neural network for distributed detection of cyber-attacks in fog-to-things communications. \cite{vinayakumar2019deep} used the DNN to develop a framework for the identification of intrusions and attacks at the network and host level. \cite{zeng2019deep} presented a lightweight framework using deep learning for encrypted traffic classification and intrusion detection. Nonetheless, little if any attention was paid to the effect of adversarial attacks against these frameworks. 

One of the first works on the vulnerability of DNN to adversarial examples was carried out by \cite{szegedy2013intriguing}. The box-constrained Limited memory approximation of Broyden-Fletcher-Goldfarb-Shanno (LBFGS) optimization algorithm was used to generate imperceptible alterations in the handwritten images in order to deceive the DNN. Although several attacks and defenses were subsequently proposed \cite{goodfellow2014explaining}\cite{kurakin2016adversarial}\cite{madry2017towards}, these attacks were designed for the computer vision field in which the vulnerability was first discovered. 

Lately, work on the effect of adversarial attacks against intrusion detection systems has been carried out. Wang \cite{wang2018deep} showed the effect of these attacks on intrusion detection systems using NSL-KDD dataset. \cite{yang2018adversarial} studied the impact of black boxes adversarial attacks on the performance of intrusion detection systems based on DNN. \cite{ibitoye2019analyzing} investigated the robustness of Self-normalizing Neural Network (SNN) against adversarial attacks on IoT networks.

According to our review of the literature, there is no work on the effectiveness of adversarial training against adversarial attacks for deep learning-based intrusion detection systems, therefore this work is presented to cover this aspect.

\subsection{Adversarial examples}
Despite the fact that deep learning has made significant progress in a variety of areas, Szegedy \textit{el al.}'s intriguing research [5] reveals that DNNs may not be as smart as they seem. They found that inserting small but carefully crafted perturbations into original images can lead to misclassification with even higher confidence. These crafted perturbations are small enough to be considered insignificant and imperceptible changes to humans.
Methods of creating adversarial examples can be categorized according to two criteria: the target class and knowledge about the model under attack.
\paragraph*{Adversarial examples' target}
given a target class $T$ different from the initial class $C^*(x) = I$ of an input $x$. An attacker seeks to find a slightly perturbed input $x'$ very similar to $x$ given a certain distance metric, yet the classifier assigns the class $C(x') = T$ to it. Thus, the \textbf{targeted adversarial attack} leads the DNN to misclassify the input as the class T desired by the attacker. As opposed to the \textbf{untargeted adversarial attack} where the objective is to find an imperceptibly modified input $x'$ so that $C^\ast (x) \neq C(x')$ which is obviously less powerful than targeted attacks. 
\paragraph*{Knowledge concerning the model under attack}
When the attacker has knowledge of everything related to the trained neural network model, including its gradients, it is a "\textbf{white box}" type attack. unlike \textbf{"black box}" type attacks where the attacker lacks knowledge of the model's gradients and has only access to the model's probability scores or, even harder, to the model's final decision. This is a common assumption for attacks on online ML services.

\section{Experimental Approach}
\label{sec:ea}

In this section, a state-of-the-art intrusion detection system based on deep learning is built to study the effectiveness of adversarial attacks. We focus on untargeted "white box" type evasion attacks, i.e., the attacker has prior knowledge of the internal architecture of the DNN used for detection and carries out his attacks during the prediction process in order to lead the system into misdetection. Subsequently, adversarial training \cite{goodfellow2014explaining} \cite{madry2017towards} is thoroughly assessed as a defense against adversarial attacks by mixing adversarial samples with clean training data during the training process to enhance the robustness of the DNN against these attacks. 

\subsection{NSL-KDD Dataset}
As one of the most commonly utilized datasets for evaluating the performance of an intrusion detection system, NSL-KDD dataset -which was released in 2009 \cite{tavallaee2009detailed}- is an enhancement of the KDD CUP'99 dataset that suffers from two major drawbacks: a huge amount of redundant records and the bias of classifiers towards frequent records. NSL-KDD addressed the two issues by removing redundant records and rebalancing the dataset classes, thereby enabling comparative analysis of different ML algorithms.

This dataset covers several attacks organized into four classes according to their nature: denial of service (DoS) attacks, probe attacks (Probe), root-to-local (R2L) attacks, and user-to-root (U2R) attacks. The records in the NSL-KDD dataset have 41 features in addition to a class label. These features are grouped into three categories: basic features, content features, and traffic features.  For the experimental part, we use KDDTrain+, which contains 125973 records, as follows: 80\% of the records are training data and 20\% are test data. Table \ref{tab:classdistro} provides a summary of the data.  

\begin{table}[!h]
	\centering
	\caption{DIFFERENT CLASSES OF THE DATASET.\label{tab:classdistro}}
	\resizebox{\columnwidth}{!}{
	\begin{tabular}{cccccc}
		\hline
		& \textbf{Normal} & \textbf{DoS}   & \textbf{Probe} & \textbf{R2L}  & \textbf{U2R} \\ \hline
		\textbf{Training data}  & 53875  & 36742 & 9325 & 796  & 42  \\ \hline
		\textbf{Test data }  & 13468   & 9185  & 2331  & 199 & 10 \\ \hline
	\end{tabular}
	}
\end{table}

\subsection{Preprocessing}
The preprocessing of the NSL-KDD dataset involves two steps: numericalization and standardization. Neural networks are unable to handle categorical values directly. Numericalization is the process of transforming these categorical values into numerical values. The features that contain categorical values in this dataset are "protocol\_type", "service" and "flag". 
Standardization is an important step to prevent the neural network from malfunctioning because of large differences between features' ranges. That is why we transform each feature into standard normal distribution.
In this paper, we focus on binary classification; therefore we qualify all attack records as "anomaly" and normal traffic as "normal". We use one-hot encoding to transform the class labels into numerical values.

\subsection{Building deep learning-based IDS}
In order to detect intrusions, a deep binary neural network with two hidden layers, each containing 512 hidden units, is implemented using TensorFlow \cite{TensorFlow}. Rectified Linear Unit (ReLU) is used as an activation function within each hidden unit so as to introduce non-linearity in these neurons' output. Following each hidden layer, a dropout layer with a dropout rate of 0.2 is employed to prevent Neural Networks from over-fitting. ADAM is set as an optimization algorithm and "categorical\_crossentropy" as a loss function to be minimized. softmax layer is added at the end to convert the logits into a normalized probability distribution. the class with the highest probability is considered as the predicted class.

\subsection{Generating adversarial samples}

We use Adversarial Robustness Toolbox (ART) \cite{nicolae2018adversarial} to implement adversarial attacks as well as the adversarial training. ART is an open-source python library for ML security developed by the International Business Machines corporation (IBM).

The generation of adversarial samples can be explained in a simple way. One can consider it as the inverse process of gradient descent where, given a fixed input data $x$ and its label $y$, the goal is to find the model parameters $\theta$ that minimize the loss function $J$. Now, to generate an adversarial sample $x'$, we proceed inversely, given fixed model parameters $\theta$, we differentiate the loss function $J$ with respect to the input data $x$ in order to find a sample $x'$ - close to $x$ - that maximizes the loss function $J$. 
FGSM \cite{goodfellow2014explaining} uses a specific factor $\epsilon$ to control the magnitude of the introduced perturbation where $\lVert x'-x\lVert< \epsilon$. The $\epsilon$ factor can be considered as the attack strength or the upper limit of the distortion amount. The adversarial sample $x'$ is then generated as follows:
\begin{equation}\label{key}
x'= x + \epsilon\nabla J_{x}(x,y,\theta)
\end{equation}
BIM \cite{kurakin2016adversarial} is another attack and is basically an iterative extension of the FGSM applying the attack repeatedly. Similar to BIM, another iterative version of the FGSM is PGD \cite{madry2017towards}. However, unlike BIM, the PGD is relaunched at each iteration of the attack from many points on the $\epsilon$-norm ball around the original input.
\subsection{Adversarial training}
\label{subsec:adv_train}
The idea behind adversarial training is to inject adversarial examples with their correct labels into the training data so that the model learns how to handle them. To do this, we use the PGD attack to generate adversarial samples before mixing them with the training data set. Here, we want to study two parameters of this defense: first, the effect of attack strength $\epsilon$ used to generate adversarial samples for the training, let's call it $\epsilon_{defense}$ to avoid confusion with the strength of adversarial attack $\epsilon_{attack}$ in the attack phase. Second, the proportion of adversarial training samples compared to clean training samples in the training data. 
\section{Experimental results}
\label{sec:er}

In this section, we first evaluate the effect of adversarial attacks on a deep learning-based intrusion detection system. then, in the second part, we examine the effectiveness of adversarial training as a means of making the system more robust against these attacks. we conclude this section by discussing and analyzing the results obtained.
\begin{figure}[htb]
	\centering
	\includegraphics[width=1\linewidth]{"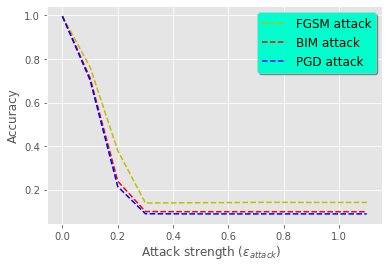"}
	\caption{Effect of adversarial attacks on deep learning-based intrusion detection system.}
	\label{fig:adv attacks}
\end{figure}

\begin{figure*}[h] 
	\centering
	\subfigure[Percentage of adversarial training samples in the training data = 30\%]{%
		\centering
		\includegraphics[width=0.5\textwidth]{"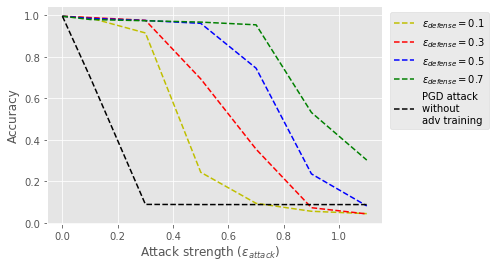"}%
		\label{fig:taux de adv samples03}%
	}\hfil
	\subfigure[Percentage of adversarial training samples in the training data = 50\%]{%
		\centering
		\includegraphics[width=0.5\textwidth]{"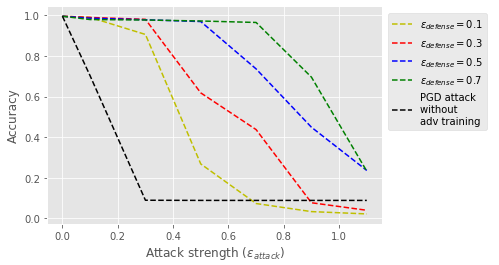"}%
		\label{fig:taux de adv samples05}%
	}
	
	\subfigure[Percentage of adversarial training samples in the training data = 70\%]{%
		\centering
		\includegraphics[width=0.5\textwidth]{"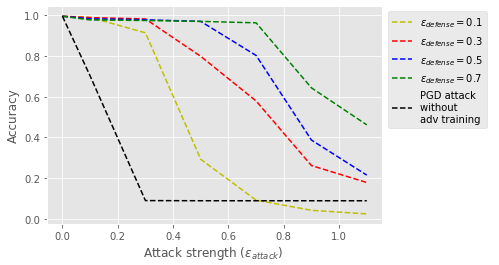"}%
		\label{fig:taux de adv samples07}%
	}\hfil
	\subfigure[Percentage of adversarial training samples in the training data = 90\%]{%
		\centering
		\includegraphics[width=0.5\textwidth]{"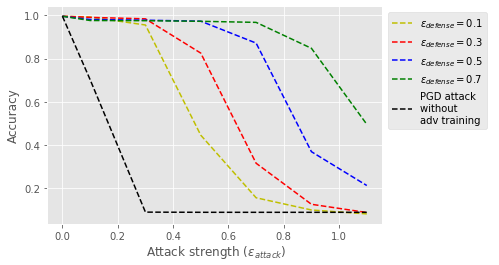"}%
		\label{fig:taux de adv samples09}%
	}
	
	\caption{Effect of adversarial training on the robustness of deep learning-based intrusion detection system}
	\label{fig:adv training}
\end{figure*}

\subsection{Effect of adversarial attacks on deep learning-based IDS }
After training our DNN model, we test its accuracy (the proportion of correct predictions among the total number of cases examined) on unmodified test data. The model gives an accuracy of 99.61\%, we then proceed to generate adversarial test data using FGSM, BIM, and PGD respectively. For each attack, the experiment is repeated, intensifying the attack by increasing $\epsilon$ value each time. Figure \ref{fig:adv attacks} shows that all three attacks deteriorate significantly the performance of the intrusion detection system. The FGSM attack lowers the accuracy of the system from 99.61\% to 14.13\%, while the BIM and PGD attacks decrease it further to 8.85\%. 

This demonstrates that, with sufficient distortion, adversarial attacks are able to defeat intrusion detection systems based on DNNs and lead them into misdetection.

\subsection{Adversarial training effect} 

As mentioned in Section \ref{subsec:adv_train}, we examine two parameters of adversarial training: 1) $\epsilon_{defense}$ which represents the attack force used to generate adversarial training samples that are mixed with clean training samples. 2) the percentage of adversarial training samples, compared to clean training samples, in the training data. Note that all the adversarial examples used are generated via the PGD attack.  

We begin by setting the percentage of adversarial training samples in the training data to 30\%, giving a fixed value of $\epsilon_{defense}$. After training the model with this mixed training data, we apply PGD attack by increasing the value of $\epsilon_{attack}$ each time. We repeat the experiment by increasing the value of $\epsilon_{defense}$ used for adversarial training as shown in Figure \ref{fig:taux de adv samples03}. The whole process is repeated by setting the percentage of adversarial training samples to 30\%, 50\%, 70\%, and 90\% respectively.

\begin{figure}[tbh!]
	\centering
	\includegraphics[width=1\linewidth]{"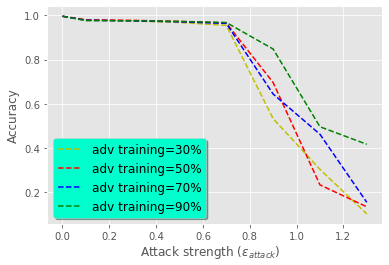"}
	\caption{Effect of the percentage of adversarial training samples in the training data, $\epsilon_{defense}=0.7$ .}
	\label{fig:ratio effect on adv train}
\end{figure}

Figure \ref{fig:adv training}  illustrates that compared to using only clean training data, adversarial training improves the robustness of the intrusion detection system against adversarial attacks. Although with sufficient attack force, the accuracy of the detector decreases considerably. We also note that increasing strength of the adversarial examples $\epsilon_{defense}$ used for the training helps to improve the robustness of the detector to some extent, making it more difficult for the attacker to create adversarial samples with a small distortion that can mislead the intrusion detection system. The same cannot be said for the impact of the percentage of adversarial training examples on the robustness of the intrusion detection system because while for $\epsilon_{defense}=0.7$, higher percentages improved the robustness of the detector against adversarial attacks as shown in Figure \ref{fig:ratio effect on adv train}, this improvement is not observed for the other values of $\epsilon_{defense}$. Thus, it is safe to say that the percentage of adversarial training examples doesn't have a direct link to the robustness of the intrusion detection system using adversarial training. This could be explained by the fact that the added dropout layers are designed to reduce overfitting effect on DNN, so as long as the model is fed with enough adversarial samples in the training phase, its performance won't change much by adding data with similar information.

\begin{figure}[tbh!]
	\centering
	\includegraphics[width=1\linewidth]{"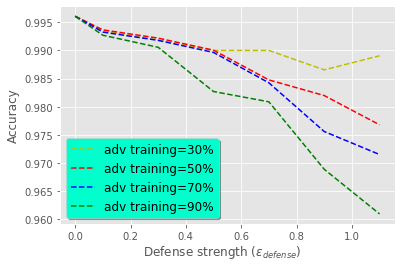"}
	\caption{Effect of adversarial training on the performance of the intrusion detection system on clean test data.}
	\label{fig:adv train on clean test}
\end{figure}

Another important aspect is the effect of adversarial training on the performance of the intrusion detection system when tested on clean test data. While results of the previous experiments indicate that adversarial training increases the robustness of deep learning-based intrusion detection systems, Figure \ref{fig:adv train on clean test} shows that adversarial training slightly decreases the accuracy of the detector when tested on clean test data. This indicates that there is a trade-off between robustness and accuracy. The decrease in accuracy of the intrusion detection system on clean test data could be explained by the fact that as the model is trained with adversarial samples, its decision boundary would change in comparison to clean data training. 

From a practical point of view, given malicious network traffic, such as HTTP traffic that wants to connect to bad URLs, such as command and control servers, the attacker can use adversarial generation techniques to transform this malicious network traffic into normal traffic for the intrusion detection system while maintaining its maliciousness, for example by adding small amounts of specially crafted data to the network traffic as padding. This allows the attacker to mislead the intrusion detection system. Adversarial training, on the other hand, is a defensive technique. It seeks to make the attacker's task more difficult by making small distortions insufficient to bypass the intrusion detection system.

\section{Conclusion and future work}
\label{sec:cfw}

In conclusion, adversarial attacks are a real threat to intrusion detection systems based on deep learning. By generating samples using adversarial attacks, an attacker can lead the system to misdetection and, given sufficient attack strength, the performance of the intrusion detection system can deteriorate significantly. As a defense against such attacks, the adversarial training was examined in depth. The results show that this method can improve to some extent the robustness of deep learning-based intrusion detection systems. However, it comes with a trade-off of slightly decreasing detector accuracy on unattacked network traffic.
An interesting future work would be to propose new defense mechanisms against adversarial attacks by exploring uncertainty handling techniques.


%
%



%
%
%

\bibliographystyle{IEEEtran}
\bibliography{bibtemplate_samples}

\end{document}